# Hindawi Mobile Information Systems

# A Smart System for Sitting Posture Detection Based on Force Sensors and Mobile Application

Slavomir Matuska, Martin Paralic and Robert Hudec

Department of Multimedia and Information-Communication Technologies, Faculty of Electrical Engineering and Information Technology, University of Zilina, Zilina/01008, Slovakia.

Slavomir Matuska: slavomir.matuska@feit.uniza.sk

## Abstract

The employees' health and wellbeing are an actual topic in our fast-moving world. The employers losing money when their employees suffer from different health problems and cannot work. The major problem is the spinal pain caused by the poor sitting posture on the office chair. This paper deals with the proposal and realization of the system for the detection of incorrect sitting positions. The smart chair has six flexible force sensors. The Internet of Things (IoT) node based on Arduino connects these sensors into the system. The system detects wrong seating positions and notifies the users. In advance, we develop a mobile application to receive those notifications. The user gets feedback about sitting posture and additional statistical data. We defined simple rules for processing the sensor data for recognizing wrong sitting postures. The data from smart chairs are collecting by a private cloud solution from QNAP and are stored in the MongoDB database. We used the Node-RED application for whole logic implementation.

## Introduction

The development of informatization currently brings new health risks. People move much less and work more often on the computer. Long-term sitting harms the spine and causes chronic problems that need long time therapy [1]. Diseased people have a significant impact on office productivity. Our motivation is to help people pay attention to their health and proper sitting in addition to work. Adopting the correct sitting position is essential for maintaining good posture and a healthy back and spine. Sitting with a straight back and shoulders will not only improve a person's physical health but can make them feel more confident. Good posture means that the parts of a person's body are correctly aligned and supported by the right amount of muscle tension.

Haynes [2], in his study, looked at the effect of sitting position on typing speed and overall well-being in people with chronic back pain. He used a unique positional wheelchair system with the possibility of position fixation and tested the efficiency of office work at variating in 6 different writing positions. The presented results showed that sitting posture has a definite impact on typing speed and user comfort.

There are several ways to monitor people's postures. Tlili [3] made an overview of systems on sitting posture monitoring. According to the principle of how to obtain this information, the systems can divide into three main categories based on:

- Computer image processing.
- Wearable clothing with sensors.
- Measuring the load distribution on some form of substrate.

Sathyanarayana [4] conducted a general review on the topic of patient monitoring based on image processing and made an overview of the algorithms used and the area of intentions that the individual systems addressed. He pointed out the limits of such systems, especially as regards the patient's distance from the camera. Obdržálek [5] used a Microsoft Kinect camera to recognize human activity and rehabilitation. His research focuses on monitoring elderly humans. He modeled the 6-different exercises and consequent positions, of which 4 of them were in sitting poses. He used the properties of Kinect to skeletonize the figure from the stereo image. Using a video system, Kuo [6] monitored the correct posture of the head against the body and spine in the sitting position. Placing the reflective markers on the human body around the head, neck, and spine simplified video signal analysis. The markers detection provides data for further estimates of the angles of curvature of the spine in the neck and cervical spine.

Systems using wearable sensors or intelligent clothing have several advantages over image-based systems. They are usually easily portable and independent of the angle of view like camera systems. The sensor is either part of the clothing or can be easily attached to the clothing. Other types of sensors can even be placed directly on human skin. Ailneni [7] used a wearable posture correction sensor to improve posture while sitting. The sensor detects postures and gives the user feedback in the form of vibrations. He claims that he can improve the posture of a man working in the office in about 25 days. The sensors are located on the head and the neck, the system reacts to incorrect head-neck position by light vibrations. Bismas [8] proposed the 3-systems for monitoring health and wellness through wearable and ambient sensors. The system is focusing on the activity monitoring of older people with incipient dementia. He designed a comprehensive sensor system that collects data from several sources located not only on the wearables but also around the living area. Analyzed data from the sensors can manage processes, e. g. automatic help calling.

Otoda [9] designed the Census sensory chair, which can classify 18 types of sitting positions. The chair has 8-accelerometers. The author states an 80% success rate of sitting position classification. Zemp [10] uses several sensors. He equipped an ordinary office chair with his custom module for motion detection. The module consists of an accelerometer, gyroscope, and magnetometer. Comparison to [9], he placed this module on the back of the chair and placed several pressure sensors on the backrest. These sensors respond by changing the resistance depending on the pressure. It is interesting in this work that they tried to analyze the measured data with various pattern recognition algorithms. They compared the following algorithms: Support Vector Machine (SVM), Multinomial Regression (MNR), Boosting, Neural Networks (NN), Radial Basis NN, Random Forest (RF) and their various configurations. The combination of 3 methods Boosting, NN, and RF has reached in this experiment the best results. Huang [11] developed a piezo-resistive sensor matrix with a thickness of 0.255 mm, which can monitor the way of sitting by a non-invasive method. The sensor field consists of two layers of polyester film.

The next chapter will deal with the concept of the whole system, how to implement it in the office room. The consequent section deals with the description of the microcontroller hardware solution implemented in the chair. In the chapter Arduino Software, we describe the concept of the application solution of the microcontroller and its communication with the server. We will also focus on the server-side based on the QIoT Suite cloud solution. QIoT features MQTT gateways, Node-RED application, and MongoDB database. In this section, we will describe carried out experiments and explain the algorithm for the detection sitting posture correctness. The last part of the system is a client application. It is a smartphone application for communication with a smart chair. It provides a fundamental interaction with the smart chair. We describe here the concept of the communication protocol between the client and the server.

## System Concept Proposal

In our system proposal, we were focusing on creating a practical smart system for sitting posture detection in the office. Our primary goal is to design a system, which could be easy to implement in any office space where the person does not have to use the same chair every day. Fig.1 illustrates the proposal of the system concept. The overall system consists of a variable number of chairs, the cloud server, and client stations - smartphones. Each chair has an electronic device based on the Arduino microcontroller, external battery power source, and six flexible force sensors. The network-attached storage from QNAP holds the cloud solution. It features the Message Queuing Telemetry Transport (MQTT) broker for communication, Node-Red for the logic, and Mongo database for data storage.

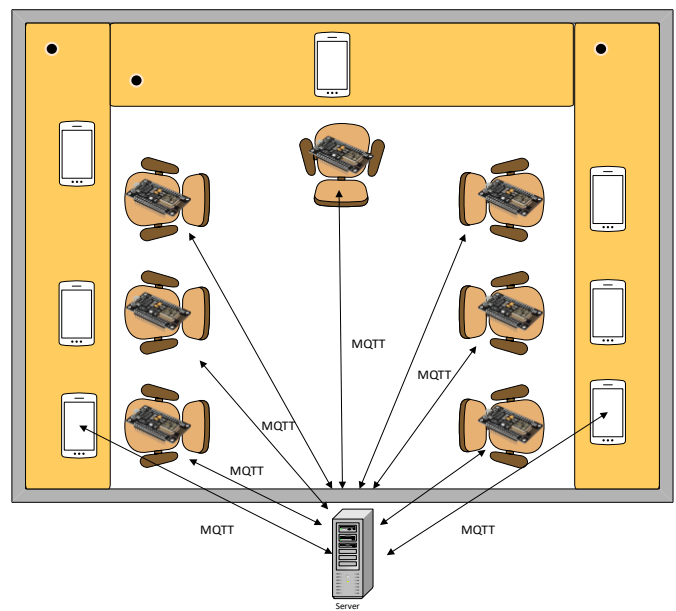

Figure 1: The system concept proposal.

The daily routine for the people working in our smart office should look like:

- The person chooses a free chair in the office and sits down. The Arduino hardware will wake up from the sleep at this point and connect to the cloud.

- The person turns on the mobile application and logs in to the chair. Each smart-chair has an identification number to login. The information about the sitting posture with additional data is displaying in the smartphone application.
- After finishing the work, the person logs out from the chair. In the final, you can view the daily report.

The following sections describe the individual parts of the system in more detail.

## Smart-Chair Hardware Design

To made a smart chair capable of measuring the pressure of the sitting person, it was necessary to embed the force sensor into the regular office chair. The six single-zone force-sensing resistors FSR402 are used to measure the force. The resistance of the sensors decreases with increasing force. Measured resistance changes from tens of kOhm to hundreds of ohms. The sensor does not need a calibration before or between the measurements. The seat is equipping with 4-sensors and the backrest with 2-sensors. The appropriate sensor positions are finding empirically. The sensors are under the lining, so they are not visible to the user. We use the NodeMcu microcontroller based on the Arduino board to collect data from sensors. The communication via Wi-Fi with the server is providing by the module ESP8266. The main advantage of this board is that it can be directly connected to the Wi-Fi and processing the data from sensors at the same time from one source code. The board supports up to 16 general-purpose input-output pins. Only one pin can work as the analog input. Fig. 2 represents the hardware schematic of NodeMcu.

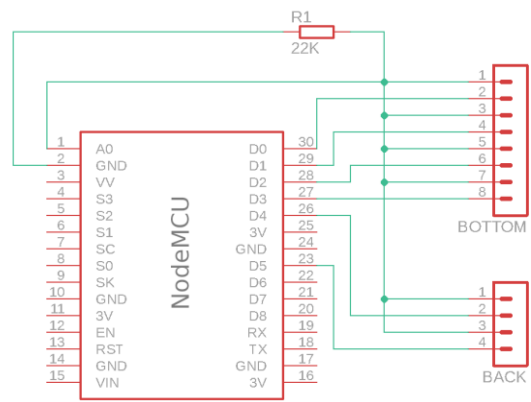

Figure 2: Hardware schematic with NodeMcu

The voltage divider with a 22 kOhm resistor measures the FSR 402 resistance. Since there is only one analog input available, we have to solve the fast switching between the sensors during the measurements. We achieved this by selecting a particular sensor with digital output from NodeMcu. The output pins from D0 to D5 were selecting the appropriate sensor.

### Arduino software

Fig.3 shows the NodeMcu flowchart diagram. The start begins with the definition of variables and their initialization. The most important variables are the WiFi network name (SSID), WiFi password, cloud server IP address, and MQTT credentials. Then the general input/output pins, serial, and WiFi communication are initializing. The first thing that the program does is connecting to the WiFi network. If the initial login fails, the system waits for 5 seconds and then retries the operation until the successful login.

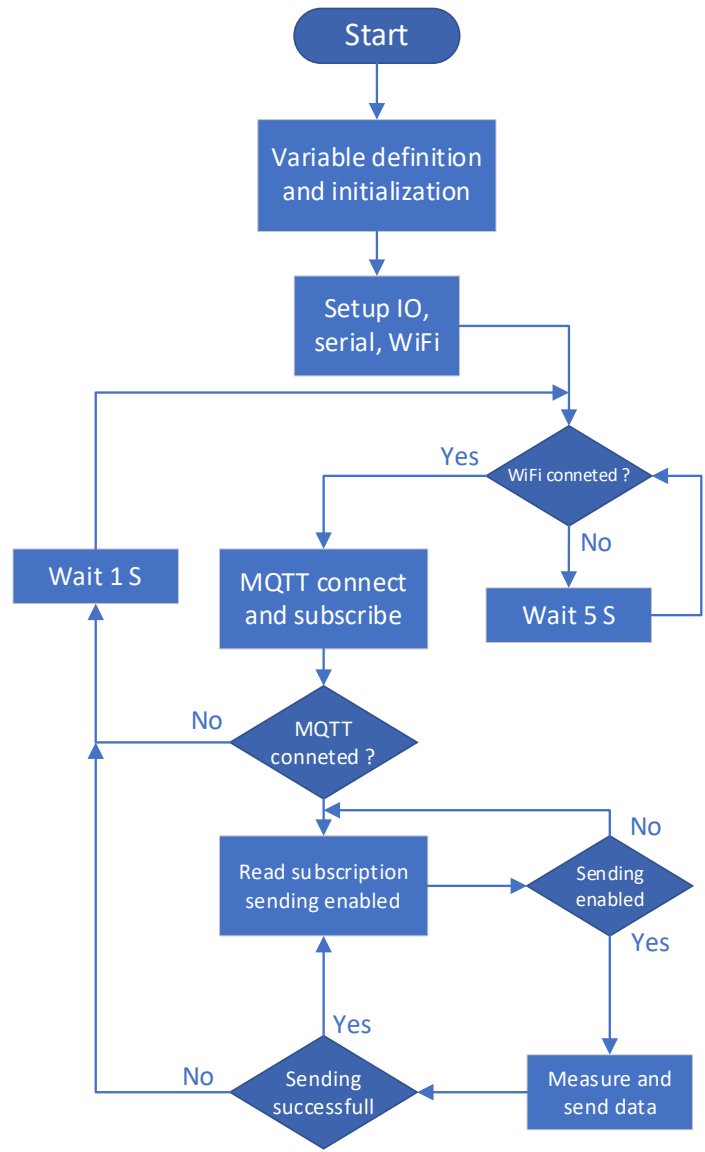

Figure 3: NodeMcu source code flowchart

The next step is building a connection with the NodeMcu. It connects to the MQTT broker using pre-defined credentials. The MQTT protocol communication is provided by an external library Adafruit MQTT Library ESP8266. To build the MQTT connection we require the client instance of the class Adafruit_MQTT_Client. This client connects to the MQTT broker. We have to create an additional object for receiving the responses from the broker. The instance of the class Adafruit_MQTT_Subscribe provides such an interface. For sending data we need an instance of a class Adafruit_MQTT_Publish. The consequence loop checks if the MQTT connection is still live. If the connection is active, the pointer to the object from class Adafruit_MQTT_Subscribe is created for fetching the new data on the subscribed channel. The channel identifier for reading commands is like:
"qiot/things/Matuska/chairs/ch" + String(CHAIR_ID) + "/sendingEnabled", where CHAIR_ID is the smart-chair identification number. On the cloud side, the MQTT broker sends the chair command using this channel. When the user connects to the smart chair using a smartphone application, the start command is issued on this channel and the NodeMcu starts the measurement and sends the data. After the user logs out, the stop command is issued from the broker. Because there is only one analog input, the resistance measurement is done in these steps:

- Set up the DX as output and toggle it to the high value.
- Read the analog value from pin A0.
- Calculate the force value from measured resistance. The calculated force represents the pressure, and its value is in the range between 0 and 15.
- Toggle to a low value and set DX as an input pin.

These steps are looping per each sensor. Afterward, the string in JavaScript Object Notation (JSON) format is prepared and published to the broker using Adafruit_MQTT_Publish object. The data are published every 1 second. If there are no active measurements incoming in time, we need to send a periodical ping command to keep the connection with MQTT alive. Without this ping, a link would break down after some time if no communication took place.

## Cloud solution and QIoT

The central unit of our smart system is the NAS from QNAP [12]. This unit runs all programs and cloud services that provide connectivity, chairs management, data storage, and data evaluation. There are two primary services:

- QIoT suite.
- Mongo DB [13].

The QIoT Suite is an application, which could be installed directly from the application center on NAS. QIoT suite integrates different services, which are necessary to provide a complex solution in the IoT world, into one application. It includes the MQTT broker [14], Node-RED, Freeboard, and supports multiple protocols and dashboards.

MongoDB is a popular, general-purpose, document-based, distributed database, which is common in a cloud solution and IoT world. All data stores MongoDB for further evaluation. Fig. 4 shows the flowchart of the system functionalities.

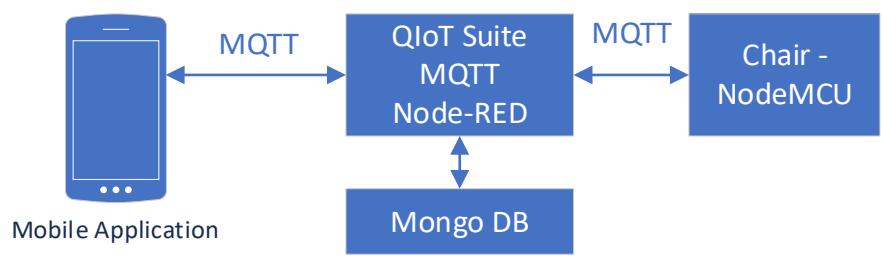

Figure 4: The system communication chain.

The smartphone application also uses the MQTT protocol to communicate with the QIoT suite server. The communication is processing via the Node-RED application. Both sides are using the MQTT communication protocol to exchange messages with the QIoT. NodeMCU sends the measured data to the cloud. Data are processed via Node-RED, stored into MongoDB, and then sent to the mobile application. More details and communication descriptions of the mobile application will be in the section Mobile application.

**QIoT Suite Lite**

QIoT Suite Lite is a complete and practical IoT private cloud platform for building and managing IoT applications. QIoT Suite leverages popular tools like Node-Red and Freebord to create an IoT environment easily and helps in efficient IoT development cases. It supports multiple protocols such as MQTT, Hypertext Transfer Protocol (HTTP), Constrained Application Protocol (CoAp). It allows us to simply create multiple dashboards and quickly connect them with multiple sensors. QIoT Suite also supports MQTT's and HTTP's security layers on the protocol for secure network connections. The suite was designed to shorten the IoT application design lifecycle and its deployment. By default, the suite provides a quick setup wizard to assist in creating IoT applications. It is also possible to use prepared codes for Python or Node.js on starting kits such as Arduino Yun, Raspberry Pi, Intel Edison, or MTK LinkIt Smart 7688. For using other kits or use a custom source code database it is necessary to create custom Thing. The QIoT home screen is shown in Fig. 5.

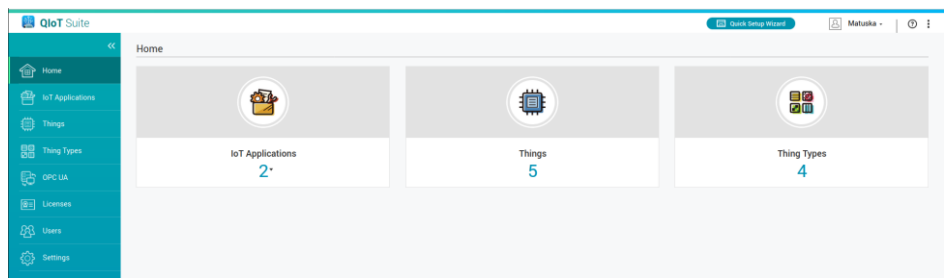

Figure 5: The QIoT home screen.

The left side of the screen provides a link to other pages. The mainframe of the screen provides links to IoT applications, Things and Things types, and their total number. The quick setup wizard is also accessible from the main screen. It is possible to create IoT applications using a quick setup wizard or custom step by step. The custom IoT application requires at the beginning only identification name. The core application is complete after the confirmation. The detail of our smart IoT system application for sitting posture detection shows Fig. 6.

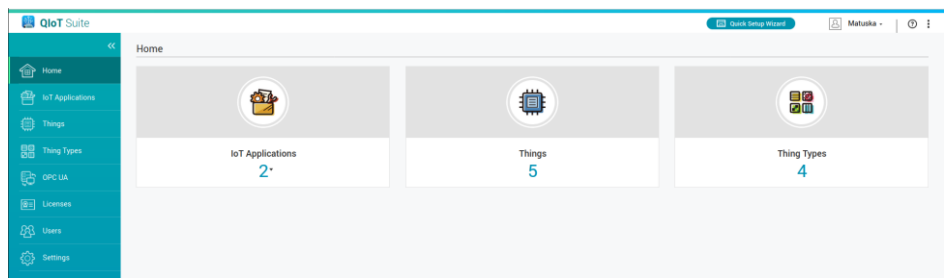

Figure 6: The detail of the IoT application.

There are three tabs on this screen, providing links to application Dashboard, Rule engine, and Things. Fig. 6. shows the table with the active Things for this application. There are two different types of Things in our application:

- General Thing type for mobile applications.
- The Chair types.

The general type is only one in the system where the mobile application uses this credential to log in to the MQTT broker. The second type is for the chairs. Each chair in the system needs to have a unique record in Things. In the QIoT suite, each Thing has its topic on the MQTT broker created automatically with the Thing creation. The topic looks like: "qiot/things/Matuska/chairDuino1", where chairDuino1 is the Thing name. The Representational State Transfer (REST) application programming interface (API) is also available for each Thing to fetch the data using the HTTP protocol. Each Thing can define one or more "resources". These resources could be sensors, peripheral, switch, or another data channel/state that needs to be transmitted or received. Adding a Resource means creating the data channel ID (MQTT → topic, HTTP, and CoAP → URL) to connect with QIoT Suite Lite. Using QioT Graphical User Interface (GUI), it is possible to generate JSON configuration file different connection types (MQTT, HTTP, or CoAP) for particular Thing. This file can be used with prepared codes to accelerate and simplify the Thing deployment. The example of this file is shown below:

```
{
    "username": "generated_username",
    "myqnapcloudHost": "Not Available",
    "clientId": "chairDuino1_1601447263",
    "host": [
        "IP address"
    ],
    "password": "generated_password",
    "port": MQTT_port,
    "resources": [
        {
            "description": "",
            "datatype": "String",
            "resourceid": "pressureData",
            "topic": "qiot/things/Matuska/chairDuino1/pressureData",
            "resourcename": "Tlakove senzory",
            "resourcetypename": "Custom Sensor(String)",
            "unit": "bar"
        }
    ]
}
```

The main advantage of the resources is that they could be implemented in a few steps as a dashboard gadget and are automatically stored in the MongoDB database if this database is configured. The resource can be imported to the dashboard using the rule engine Qbroker. There are two methods to use Qbroker:

- Importing data from the resource.
- Importing data from the rule engine.

After a few steps, the resource value will appear on the dashboard. The dashboard supports different kinds of widgets, such as text, gauge, sparkline, pointer, indicator, action widget, or QioT historic chart. The third tab is the Rule engine based on Node-RED and features a flow editor to simplify IoT development. There are four customized QIoT nodes:

- QDashboard – Provides a live data API endpoint.
- QBroker(in) – Receives Thing data.
- QBroker(out) – Transmits Thing data.

QHistoricData - retrieves the maximum, minimum, and average values stored in the database for defined resources.

**Node-RED application**

The Node-RED application implements the logic part of the smart system for sitting posture detection on the private cloud platform. Such an application is usually splitting into the program Flows. There are two main flows in our solution:

- The flow for pressure data processing.
- The flow for chairs and mobile application management.

Fig. 7. shows the schematic design of flow for chairs and mobile application management.

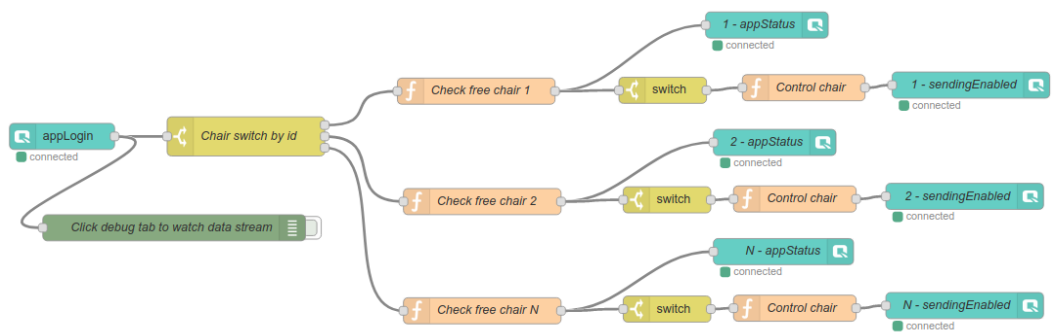

Figure 7: Node-RED flow for chairs and mobile application management.

The flow starts with the QBroker in node. This input node listens on the topic "appLogin". The purpose of this flow is to manage chairs and mobile applications and there is only one common input node for all chairs and applications in this flow. All data coming through the Qbroker node have to be in JSON format. The example of the expected JSON response follows data:

```
{
        "chairId": 1,
        "query":"login"
}
```

The chairId specifies the identification number of the chair. The tag "query" represents the performed action. The action could be "login" or "logout" action to the chair. The message is routed in the switch node depending on the chair identification number "chairId". Then the system checks if the selected chair is unoccupied. The function for checking the chair state adds an attribute to the message about the operation success. The message with the response is then sent to the mobile application via the Qbroker out node on the topic "chairs/ch1/appStatus". The message is also router via another switch, which evaluates if the added attribute is true or false. If the added attribute is true, the command message is sent to the chair using the topic "chairs/ch1/sendingEnabled". The sent message contains a command to start or stop sending the pressure data. There are two QBroker out nodes for each chair. If there is a request for adding a new chair to the system, it is necessary to copy the whole block for the chair and change the topic name according to the chair identification number. This could be done easily with a few steps. Fig. 8 shows the flow for pressure data processing.

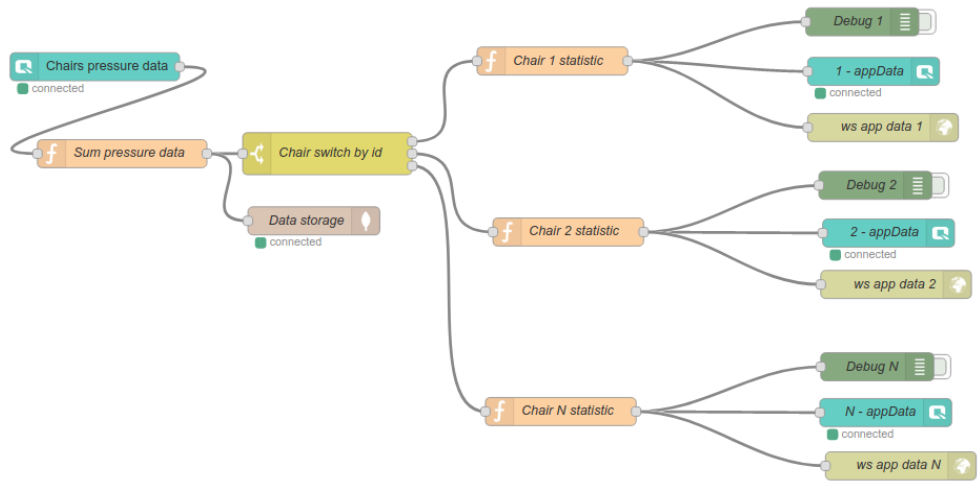

Figure 8: Node-RED flow for pressure data processing.

The flow starts with the QBroker in the node. This input node listens on the topic "chairPressureData." The purpose of this flow is to collect, evaluate, and propagate the chair pressure data. There is only one common input node for all chairs in this flow. Only chairs publish the data on this topic. The example of sending data by "publish" command is:

```
{
        "chairId":1,
        "data": ["6.04", "6.21", "7.80", "6.75", "2.21", "1.35"]
}
```

The next node "Sum pressure data" sum the pressure data and add the message timestamp in Unix format. The message is routed in the switch node based on the chair identification. The next node "Chair ID statistic" is the most important in the system. In this node, the function implements the whole logic for bad sitting posture detection. The Qbroker out is implemented for each chair with the topic "chairs/ch1/appData", where ch1 is chair identification number. The WebSocket node is also implemented with URL "/ws/ch1/appData". The same message is sent via MQTT Qbroker and WebSocket. Websocket is implementing because we expect further communication with the client application. The sent message structure is:

```
{
        "chairId":1,
        "data":["5.63","5.70","5.61","5.51","2.64","5.71"],
        "sum":30.80,
        "actual_time":1601455127,
```

```
            "avg":5.612500000000001,
            "deviation":0.00461875,
            "chdata":{
                "actual_sitting_state":"green",
                "avg_deviation":0.0070274999999999825,
                "avg_back_deviation":2.790704999999999,
                "chair_id":1,
                "actual_sitting_time":45,
                "back_data_present":1,
                "long_sitting":0,
                "duration":1358,
                "start_time":1601453768.362,
                "sitting_history":[
                    {"timestamp":1601455120.569,"sitting_status":1},
                    {"timestamp":1601456131.239,"sitting_status":0},
                    {"timestamp":1601456822.133,"sitting_status":1}],
                "actual_sitting_status":1
                }
}
```

where "chairId" is the chair identification number, "data" contains the pressure values from sensors, "sum" is the sum of the pressure values and "actual_time" is the message timestamp. The "avg" stands for actual average and "deviation" is the actual standard deviation of values from four bottom sensors. JSON object "chdata" contains information about sitting posture for the mobile application. Attributes meaning is explained in Table 1.

Table 1: Description of chair data attributes.

| Attribute | Explanation | Type/States |
|---|---|---|
| actual_sitting_state | Information about sitting posture. | String {green, orange, red} |
| avg_deviation | Average standard deviation from the last five measurements for four bottom sensors. | Float |
| avg_back_deviation | Average standard deviation from the last five measurements for two back sensors. | Float |
| actual_sitting_time | Current uninterrupted sitting on the chair in seconds. | Integer |
| back_data_present | Indicates if there are present valid data from back sensors. | Integer {0, 1} |
| long_sitting | Indicates long uninterrupted sitting on the chair. | Integer {0, 1} |
| duration | The elapsed time in second from the login. | Integer |
| start_time | Login time. | Unix timestamp |
| sitting_history | History of the sitting states with the Unix timestamp, when the changes occurred. | Array of objects |
| actual_sitting_status | Indicates if there is somebody sitting in the chair. | Integer {0, 1} |

Object "chdata" is added to the message in the node "Chair 1 statistic". Fig. 9 describes the function node flowchart.

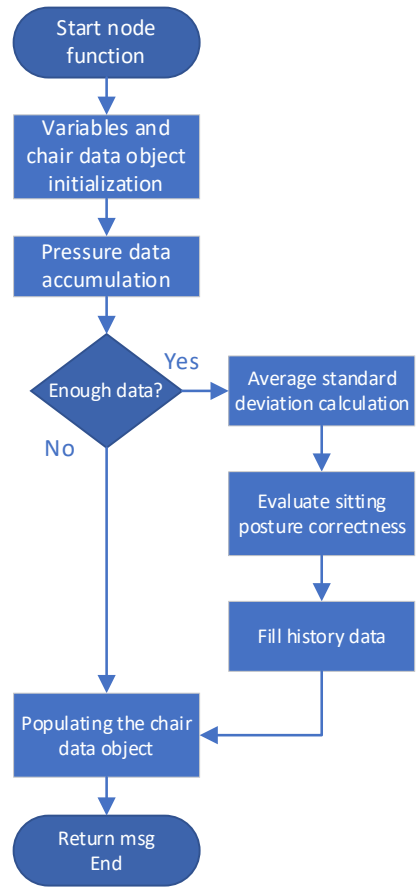

Figure 9: Sitting posture evaluation flowchart.

The function starts with variable initialization. If this function is executed for the first time after the user was logged in, the object "chdata" is created with initial values. The object contains all necessary variables for statistical computations and information about sitting posture propagated to the mobile application. The function accumulates the pressure data from the chair. The function starts to recalculate statistics as soon as it collects data in the last 10 measurements. The last 10-measurements are using to detect a seated person. The longer time is suppressing the false detection of a state change of sitting or standing. For the computation of the standard deviation, only the last 5-measurements are using. Deviations are calculated separately for back sensors and separately for seat sensors. These deviations are the primary features of the incorrect seating position detection algorithm.

Table 2: Rules for the incorrect seating detection algorithm.

| Sitting state | Rules |
|---|---|
| Green | avg_deviation < ODT && back data presented |
| Orange | (avg_deviation > ODT && avg_deviation < RDT)  or (avg_deviation > OCDT && avg_deviation < ODT && back data NOT presented) |
| Red | (avg_deviation > RDT) or (avg_deviation > ODT && back data NOT presented) |

For the sitting poses detection by the proposed method, it was necessary to determine empirically 3-threshold values of deviation. These threshold deviations are computing from the sensors in the seating area. In the seating part, we use Orange Deviation Threshold ODT=3.0, Red Deviation Threshold RDT=6.8, and Orange Conditional Deviation Threshold OCDT=0.8. A summary of the rules is given in Table 2.

Based on our previous work [15] and the findings presented in [16] and [17], we defined 9-different sitting posture for the further examination. Fig. 10 shows the defined sitting postures.

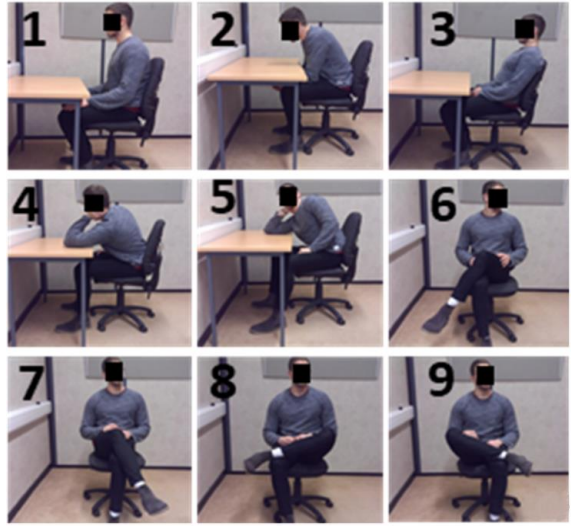

Figure 10: Tested sitting posture [15].

The sitting position number 1 is considering as a correct sitting posture. The positions 2-5 represent the orange state, and 6 - 9 is the red state. To find out the threshold values, we carried out the experiments to measure the standard deviation for each posture. Twelve test subjects participated in this experiment.

Table 3: Average standard deviation for each sitting posture.

| Posture/Subject | 1 | 2 | 3 | 4 | 5 | 6 | 7 | 8 | 9 |
|---|---|---|---|---|---|---|---|---|---|
| 1 | 0.06 | 1.2 | 1.15 | 7.2 | 6.2 | 12.3 | 13.31 | 11.03 | 10.45 |
| 2 | 0.2 | 0.4 | 0.24 | 3.62 | 2.27 | 14.03 | 12.92 | 12.43 | 12.3 |
| 3 | 0.05 | 0.6 | 4.53 | 1.83 | 1.3 | 9.06 | 8.29 | 11.83 | 12.72 |
| 4 | 0.14 | 0.76 | 1.78 | 0.79 | 2.12 | 4.02 | 11.23 | 0.65 | 1.53 |
| 5 | 0.19 | 0.54 | 8.42 | 1.74 | 1.27 | 8.9 | 9.4 | 10.84 | 11.87 |
| 6 | 0.06 | 0.23 | 0.32 | 1.56 | 3.13 | 8.46 | 4.81 | 1.76 | 2.5 |
| 7 | 0.24 | 0.57 | 4.15 | 0.95 | 1.73 | 9.37 | 1.49 | 4.96 | 7.01 |
| 8 | 0.37 | 1.99 | 9.95 | 15.93 | 11.65 | 11.15 | 7.96 | 8.99 | 9.61 |
| 9 | 0.3 | 0.87 | 0.64 | 10.48 | 6.41 | 13.02 | 9.53 | 9.52 | 11.08 |
| 10 | 0.28 | 0.86 | 13.52 | 4.72 | 2.3 | 13.6 | 1.87 | 12.39 | 13.14 |
| 11 | 1.27 | 1.2 | 0.18 | 4.25 | 0.96 | 9.84 | 0.8 | 5.93 | 0.75 |
| 12 | 0.03 | 0.98 | 0.21 | 1.84 | 1.83 | 1.12 | 0.77 | 2.04 | 2.55 |
| Average | 0.266 | 0.850 | 3.758 | 4.576 | 3.431 | 9.573 | 6.865 | 7.698 | 7.959 |
| STDEV | 0.335 | 0.467 | 4.534 | 4.586 | 3.152 | 3.840 | 4.729 | 4.421 | 4.820 |

For each subject, we recorded 10-measurements for each posture. Then we calculated the average standard deviation for each pose. The next step was the average calculation for each posture. Afterward, we determined the 3-threshold values based on the collected data. The Table 3 contains all obtained and calculated data.

When new data arrives, unnecessary data is released using the first in-first out method FIFO. If at least one amount exceeds the threshold in a series of 10 measurements, it is considering as a continuous seating. The mechanism for detecting the presence of a seated person seeks to eliminate false detections of leaving the chair. The feature to compute is the sum of pressures from each sensor. If this sum(Si)< 1 for 10-consecutive measurements, leaving the chair is detected. Thus, short standing up or reaching for the object will not be considered as leaving the chair. The system cumulatively calculates the continuous sitting time and, after exceeding the time threshold, which is currently set to 1 hour, sends the flag for long sessions and changes the status to red.

## Mobile Application

The smartphone application serves as client access to the smart chair measurements. In the first step, the user must connect to the MQTT server. The user should fill the connection form with the server address, communication port, user login, password, and chair identification number. The login activity is in Fig. 11 with filled connection data.

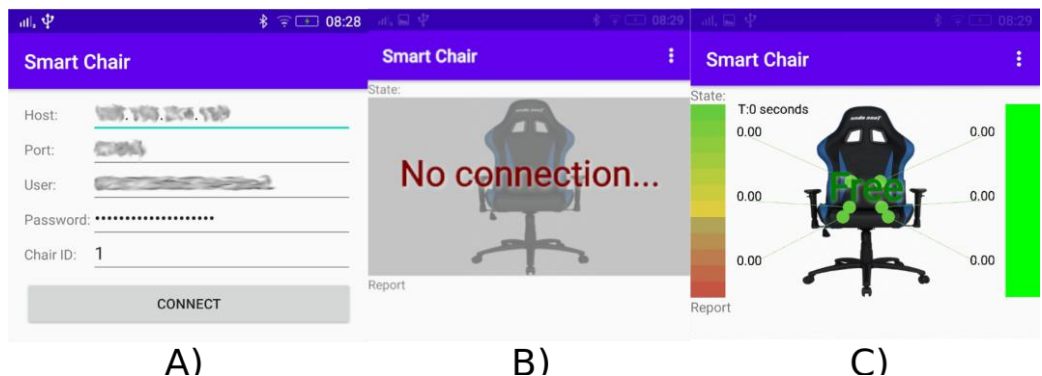

Figure 11: The connection to the MQTT server and login to the chair – a) Login activity, b) Succeed connection to MQTT server, c) Succeed logging to chair.

The data from multiple chairs store on one server. The successful connection to the MQTT server is followed by automatic login to the selected chair. If the Smart Chair is occupying another person or chair is inactive at the time. In such a case application is connected to the MQTT server but has no connection to the chair. Fig. 11b is displaying the state when the MQTT link is open, but the chair is still not accessible. The current active client must release the chair to make further connections. The client application allows only one active connection per chair. If login to Smart Chair is successful, the full chair image appears with its measured parameters. The current sitting position is representing by the colors. Additional sound alerts signalize poor sitting postures. The color representation is:

- Green - The client is sitting in the correct position with an evenly distributed load. The Unoccupied Smart Chair is also green, with an additional title "Free," as shown in Fig. 11c.
- Orange - The participant is sitting, but his weight is not distributing evenly.

- Red - The participant is sitting in an unhealthy sitting position. This position is signalizing an excessive load on one side. This condition may also occur when the client sits continuously for more than an hour.

The first color scale panel to the left of the chair represents the measuring range, in Fig. 12, where the green color is the lowest pressure, and the red is the highest pressure. The second single color panel to the right of the chair picture represents the basic sitting state {green, orange, or red}. The circles represent sensor positions, and their color depends on the pressure, following the color palette. In the middle is a symbolic representation of a chair. The numbers (no units) are pressure measurements that express the intensity of the load. The color of individual points is adjusting according to the current load. Fig.12a represents an evenly distributed body weight on the pressure sensors following the orange state. After transferring the scale to one side of the chair, the system detects the incorrect position of Fig. 12b seating condition (condition orange). In the orange state, the application plays a short quiet tone ("elevator tone"). It unobtrusively alerts the user to the wrong sitting position. In case when the user load distribution is extreme, the state changes to the red state. The red state is signalizing by an annoying audio signal (alert tone). The user has the option to turn the application of or move his body to a healthy sitting position to stop the annoying alarm. The bottom of the screen reports different events. The report can hold a history of sitting states, variety events, logging and connection issues, debug information.

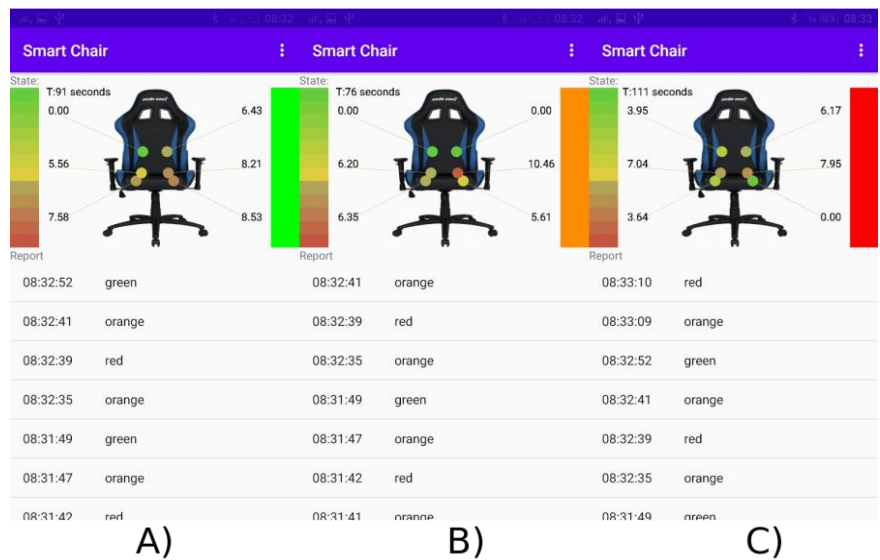

Figure 12: The examples of sitting states a) green (standard sitting), b) orange (bad sitting), c) red (heavy load for the backbone).

Fig. 13 shows the schematics of the mobile application and its communication with the MQTT server. The design of modules for communication with the MQTT server is considering to be a reusable code for further applications. The login activity is the first screen that the user can see. He passes the login data to the MqttChair communication module. This module uses the MQTT communication protocol services with the help of the "eclipse / paho.mqtt.android" library. The inputs to the module are the login data to connect the server and the necessary user and chair identification. First, it ensures the establishment of a connection with the MQTT server. The entire communication takes place asynchronously since it is event-controlled. After successfully connecting to the server, it subscribes to the subscription, whether the chair is available SUBSCRIBE_LOGIN. Subsequently, a request to join the PUBLISH_LOGIN chair is published. Availability LOGIN_ACK returns information about whether the Smart Chair is occupying by another participant. If the Smart Chair is unoccupied, the module MqttChair subscribes to collecting chair data using SUBSCRIBE_DATA. New data is receiving via the call-back function.

The communication ends when the user requests a logout action, or the connection is unexpectedly broken. The standard way is to send the message PUBLISH_LOGOUT and consequently disconnect from the MQTT server. The response to logout action is an acknowledgment of LOGOUT_ACK. The communication with the server is user terminated by sending a message DISCONNECT and subsequently confirmed by DISCONNECT_ACK. The MQTT module provides an interface with two listener call-back functions. The first function onReport() returns a message designed to monitor traffic and generate events. The second function onDataReceived() provides data received from the smart chair. To transfer chair data internally the ChairData object is used. The most user interaction is performed by the Monitor Activity. This activity represents the main GUI for the user, sends commands to the MqttChair module, receives data and messages using MqttChair listeners. It provides basic functionality for user login and logout actions. The Monitor Activity is also holder place for the ChairView widget. The ChairView widget is a module for displaying Smart Chair data. The input is the instance of the ChairData object. The document-view architecture system is used, where the document is represented by the ChairData object, and the view is displayed by the ChairView module.

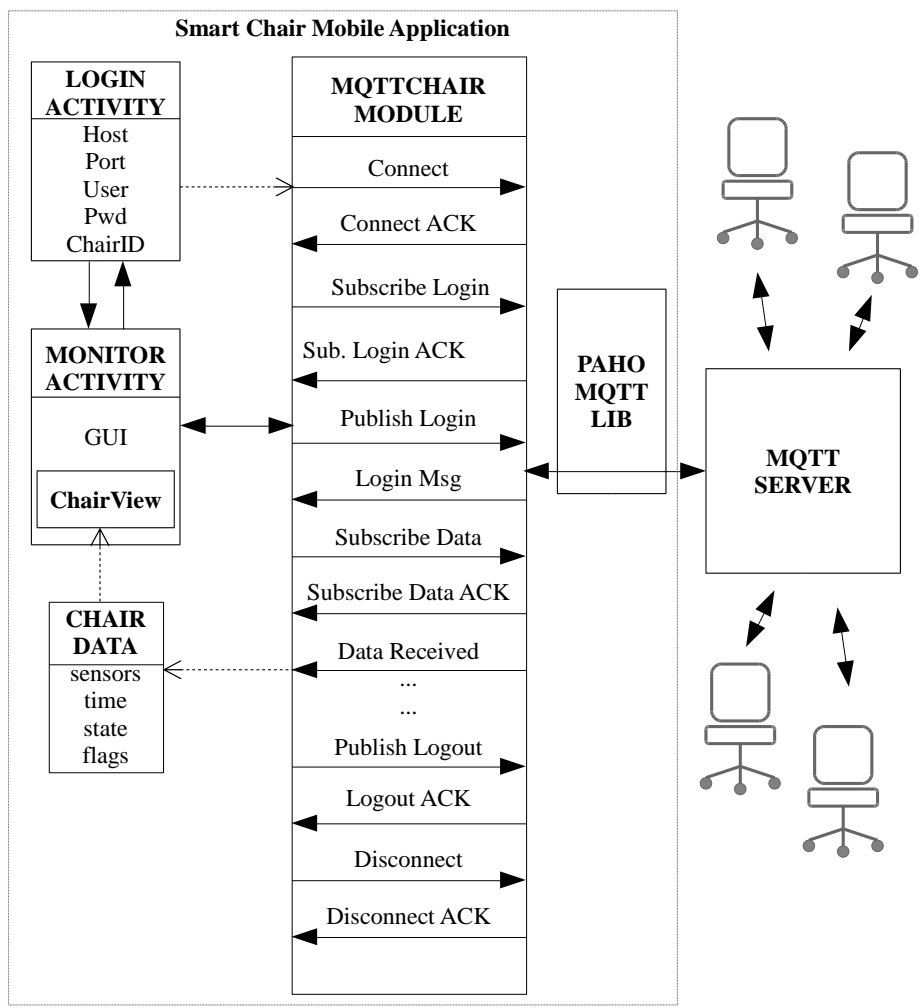

Figure 13: The communication between the mobile application and MQTT server.

## Discussion

The most crucial part of our proposed smart chair is the hardware uptime without the need for a recharge. During the testing, we used the external power bank with a capacity of 4000 mAh as the power source. With this external source, our hardware can run up to 48 h of active measurement. For most of the day, the chair is not occupied and the NodeMCU could enter the sleep mode and wake up only when someone is sitting on the chair. This will theoretically improve uptime up to 12 days and thus can be used in a real application. The other issue is implementing the force sensors into the chair. This process is rather difficult and the implementation itself takes some time. On the other hand, if we want to produce the smart chair in numbers, it will be necessary to create a more automated way. Software deployment on the server-side is easy and straightforward and while it can run on as cheap hardware as Rasberry Pi, it is also cost-effective. Overall, the proposed system is easy to implement. The user receives the notifications about the sitting posture correctness on the mobile application on Android. For the Apple users, it is necessary to develop an application for iOS or use the multiplatform framework to write one application for both platforms with a common codebase. Using the information from the application users can adjust their sitting customs and improve their health and wellbeing easily.

## Conclusions

This paper presents a smart IoT system for sitting posture detection based on force sensors and mobile applications. Six flexible force sensors, two on the backrest and four on the bottom seat, were embedded in the office chair. NodeMCU board was used to measure the sensor's resistance and sends the data to the cloud using the MQTT protocol. The data are stored and evaluated on the cloud using Node-RED and MongoDB. The user can see the information about sitting posture correctness and other detailed information in the mobile application. Our goal was to create simple rules to detect correct sitting posture in the term of minimal computation power requirements. We defined 9-sitting postures and carried out the experiments on 12 people to identify the rules. The standard deviation from the bottom seat sensors shows a strong dependence on sitting postures. The calculation of the standard deviation is not hard for computing power. Based on our observation, we divided the sitting posture correctness into three groups, namely green, orange, and red, and defined the three threshold values. We used these threshold values for the evaluation of the sitting posture into one of the three groups. The system also calculates the time of sitting without a break and informs the user in the case of long-term continuous sitting. We deploy our solution on the network-attached storage from QNAP. But because of the evaluation algorithm simplicity, it is possible to run our smart system on the cheap hardware such as Raspberry Pi with only minimal changes. The QIoT specific gateway must replace it with a regular MQTT gateway. It is also necessary to install services like Mosquito MQTT broker, standalone Node-RED application, and optional MongoDB manually on Raspberry Pi.

### Data Availability

The source codes are available on the request from the corresponding author.

### Conflicts of Interest

The authors declare that they do not have any conflicts of interest.

## Acknowledgments

The work presented in the paper was supported by the European Union's Horizon 2020 research and innovation program under the Marie Skłodowska-Curie grant agreement No 734331 and by the project "Competence Centre for research and development in the field of diagnostics and therapy of oncological diseases", ITMS: 26220220153, co-funded from EU sources and the European Regional Development Fund.